\documentclass[secnumarabic,amsmath,amsfonts,amssymb, notitlepage,floatfix, nofootinbib, aps,reprint,prd]{revtex4-1}
\usepackage[utf8]{inputenc}
\usepackage{graphicx}

\begin{document}
\author{Hilario Pérez}
\email{hilario.pr@hotmail.com}
\author{Piotr Kielanowski}
\email{kiel@fis.cinvestav.mx}
\affiliation{Departamento de Física, Centro de Investigación y de Estudios Avanzoados del IPN, México D.F., Mexico}
\author{S. Rebeca Juárez W.}
\email{rebecajw@gmail.com}
\affiliation{Departamento de Física, Escuela Superior de
  Física y Matemáticas, Instituto Politécnico Nacional,
  U.P. ``Adolfo López Mateos'' C.P.~07738, México D.F., Mexico}
\author{Gerardo Mora}
\email{gerardo.mora@ujat.mx}
\affiliation{División Académica de Ciencias Básicas, Universidad ``Juárez'' Autónoma de Tabasco, Apartado Postal 24, Cunduacán Tabasco 86690, Mexico}
\title{Asymptotic properties of CP violation in the Standard Model}

\begin{abstract}
We present the analysis of the renormalization based evolution of the CP-violation observables obtained from the $C$ matrix introduced by
Jarlskog. We show that the observables $\vert\det C\vert$ and $\operatorname{Tr}C^{2}$ decrease very fast with the energy and their value
is reduced at the Planck's scale by~5 and~3 orders of magnitude with respect to their low energy values. On the other hand the Jarlskog's
CKM matrix rephasing invariant $J$ increases with energy and at the Planck's scale is 25~\% larger than at low energy. The absolute value
of the coefficient $a_{CP}\sim \det C/(\operatorname{Tr}C^{2})^{3/2}$ decreases with energy and at the Planck's scale it is 12~\% smaller than at low energy. We also find that the pattern of the eigenvalues of the $C$ matrix is such that two eigenvalues almost cancel each other and their absolute values are much bigger than the absolute value of the third eigenvalue. The low rate of the CP-violation is a consequence of this pattern of the eigenvalues.
\end{abstract}
\maketitle
\section{Introduction}\label{sec:introduction}
The discovery of the Higgs boson and the measurement of its mass~\cite{Aad20121,*Chatrchyan201230} determine the last unknown parameter of the Standard Model~\cite{Glashow1961579,*salam,*PhysRevLett.19.1264}. In such a way the Standard Model becomes a fully predictive theory~\footnote{This is not true for extensions of the Standard Model, which usually contain additional unknown parameters. The known value of the Higgs mass can, at best, put some constrains on these additional parameters.}. In this paper we study the renormalization group evolution of the CP-violation parameters in the Standard Model.

The CP-violation in the Standard Model has its origin in the complex values of the quark Yukawa couplings, which result in a complex Cabibbo-Kobayashi-Maskawa matrix, with a phase, which cannot be eliminated by the quark rephasing freedom of the Standard Model~\cite{PhysRevLett.10.531,*Kobayashi1973}. The condition for the presence of the CP-violation in the Standard Model in terms of the mass matrices (or quark Yukawa couplings) has been given by Jarlskog~\cite{PhysRevLett.55.1039,*Jarlskog1985}. She considers the commutator of the quark mass matrices 
\begin{equation}
\label{eq:1}
i\tilde{C}=[M,M'],\quad \det\tilde{C}\neq0 \text{ for CP violation}
\end{equation}
($M$ and $M'$ are the mass matrices of the up- and down-quarks, respectively) and shows that a non vanishing $\det\tilde{C}$ signifies the presence of CP-violation in the Standard Model. Based on this analysis we consider the renormalization group evolution of all parameters that describe the properties of CP-violation in the Standard Model. The first renormalization group analysis of $\det C$~\cite{PhysRevLett.57.1982} was performed almost 30 years ago and it was done with one loop equations. The first study of the evolution of $J$ was done almost 25 years ago~\cite{Jin1990101} and it was based on the renormalization group equations for the absolute values of the CKM matrix elements. Our analysis of CP-violation is based on the two loops equations for the quark Yukawa couplings and it is the first complete analysis based on the Jarlskog's matrix $C$.

In Section~\ref{sec:RGE} we briefly recapitulate the Renormalization Group Equations (RGE) in the Standard Model. Section~\ref{sec:Jarlskog} is devoted to the discussion of the Jarlskog's analysis of CP-violation and in Section~\ref{sec:CP-evolution} we discuss the renormalization group equations for the CP-violation observables. Section~\ref{sec:discussion} contains the main results of the paper and in Section~\ref{sec:conclusions} we give a general view of the obtained results.

\section{Renormalization Group Equations}\label{sec:RGE}

Renormalization Group (RG) analysis in field theory is the most important method for the asymptotic analysis of the theory at high energies~\cite{Stueckelberg1953,*PhysRev.95.1300}. In the Standard Model (SM) the Renormalization Group Equations (RGE) were used to find the behavior of gauge couplings at high energies (asymptotic freedom~\cite{PhysRevLett.30.1343,*PhysRevLett.30.1346}), to justify the grand unified extensions of the SM~\cite{PhysRevD.10.275,*PhysRevLett.32.438} and also to obtain limits on the Higgs boson mass and for the determination of the range of validity of the SM, imposing the conditions of triviality and stability of the model (see~\cite{PhysRevD.72.096003} and references therein). With the complete knowledge of the parameters of the~SM one can now give precise answers about the energy evolution of the model.

The generic form of the RGE equation for an observable~$x$ is the following
\begin{equation}
\label{eq:2}
\frac{dx }{dt}=\beta_{x},
\end{equation}
where $t$ is the renormalization point energy in suitable units (we use $t=\ln(E/m_{t})$ and $m_{t}$ is the top quark mass) and $\beta_{x}$ is the \textit{beta} function of the parameter~$x$ and it has the generic form of a perturbative series
\begin{equation}
\label{eq:3}
\beta_{x}=\frac{1}{16\pi^{2}}\beta_{x}^{(1)}+ \frac{1}{(16\pi^{2})^{2}}\beta_{x}^{(2)}+\cdots
\end{equation}
Here $ \beta_{x}^{(i)} $ are the $i$~loops contributions to the $\beta_{x}$ function. In the SM the $\beta_{x}^{(i)}$ functions are fully known up to two loops~\cite{PhysRevLett.90.011601}. There are some partial results with more loops, but we do not include them, since they are not complete, so they do not improve the precision of the analysis.

In the SM the full set of parameters $x$ of the RG evolution is given in Table~\ref{table:1}.
\begin{table}[h]
\centering
\begin{tabular}{r|l}
$x$&Description\\
\hline
$g_{1},g_{2},g_{3}$&gauge couplings\\
$Y_{u},Y_{d}$&quark Yukawa couplings matrices\\
$Y_{e}$&lepton Yukawa couplings matrix\\
$m^{2},\lambda$&parameters of the Higgs scalar potential,\\
&$\lambda$ is the Higgs quartic coupling
\end{tabular}
\caption{Parameters of the SM}
\label{table:1}
\end{table}
The RGE for $g_{1},g_{2},g_{3},Y_{u},Y_{d},Y_{e}$ and $\lambda$ do not depend on $m^{2}$, so we do not have to consider the RGE for $m^{2}$.

The one loop $\beta_{x}^{(1)}$ functions in the SM are equal to
\begin{equation}
\label{eq:4}
\begin{split}
\beta_{g_{1}}=&\frac{41}{10}g_{1}^{3},\quad
\beta_{g_{2}}=-\frac{19}{6}g_{2}^{3},\quad
\beta_{g_{3}}=-7g_{3}^{3},\\
\beta_{Y_{u}}=&Y_{u}\left(\frac{3}{2}(Y_{u}^{\dagger}Y_{u}-Y_{d}^{\dagger}Y_{d})+Y_{2}(S)\vphantom{-(\frac{17}{20}g_{1}^{2}+\frac{9}{4}g_{2}^{2}+8g_{3}^{2})}\right.\\
&\hspace*{0.3\linewidth}\vphantom{\beta_{Y_{u}}=Y_{u}}\left.\vphantom{\frac{3}{2}(Y_{u}^{\dagger}Y_{u}-Y_{d}^{\dagger}Y_{d})+Y_{2}(S)}-(\frac{17}{20}g_{1}^{2}+\frac{9}{4}g_{2}^{2}+8g_{3}^{2})\right),\\
\beta_{Y_{d}}=&Y_{d}\left(\frac{3}{2}(Y_{d}^{\dagger}Y_{d}-Y_{u}^{\dagger}Y_{u})+Y_{2}(S)\vphantom{-(\frac{1}{4}g_{1}^{2}+\frac{9}{4}g_{2}^{2}+8g_{3}^{2})}\right.\\
&\hspace*{0.3\linewidth}\vphantom{\beta_{Y_{d}}=Y_{d}}\left.\vphantom{\frac{3}{2}(Y_{d}^{\dagger}Y_{d}-Y_{u}^{\dagger}Y_{u})+Y_{2}(S)}-(\frac{1}{4}g_{1}^{2}+\frac{9}{4}g_{2}^{2}+8g_{3}^{2})\right),\\
\beta_{Y_{e}}=&Y_{e}\left(\frac{3}{2}Y_{e}^{\dagger}Y_{e}+Y_{2}(S)-\frac{9}{4}(g_{1}^{2}+g_{2}^{2})\right),\\
\beta_{\lambda}=&12 \lambda^2-\left(\frac{9}{5} g_{1}^{2}+9 g_{2}^{2}\right) \lambda\\
&\hspace*{0.2\linewidth}+\left(\frac{27}{100} g_{1}^{4}+\frac{9}{10} g_{1}^{2}
   g_{2}^{2}+\frac{9}{4} g_{2}^{4}\right)\\
   &\hspace*{0.4\linewidth}+4 \lambda Y_{2}(S)-4 H(S).
\end{split}
\end{equation}   
Here $Y_{2}(S)$ and $H(S)$ are auxiliary functions equal to
\begin{equation*}
\begin{split}
   &Y_{2}(S)=\operatorname{Tr}(3Y_{u}^{\dagger}Y_{u}+3Y_{d}^{\dagger}Y_{d}+Y_{e}^{\dagger}Y_{e}),\\
   &H(S)=\operatorname{Tr}(3(Y_{u}^{\dagger}Y_{u})^{2}+3(Y_{d}^{\dagger}Y_{d})^{2}+(Y_{e}^{\dagger}Y_{e})^{2}).
\end{split}
\end{equation*}
The explicit form of the two loop beta functions is rather long, so we do not repeat them here, but we use the two loop beta functions from Ref.~\cite{PhysRevLett.90.011601} .

From Eqs.~\eqref{eq:4} one can see that the one loop RGE for $g_{1},g_{2},g_{3},Y_{u},Y_{d},Y_{e}$ do not depend on the quartic coupling constant $\lambda$, so they are insensitive to the Higgs mass. This is not the case for the two loop equations. This is the reason for which the complete RG analysis of the SM should be performed at a higher level than the one loop approximation.

In Section~\ref{sec:CP-evolution} we will use Eqs.~\eqref{eq:4} and the two loop RGE for the determination of the evolution of the CP-violating observables.

\section{Jarlskog's Description of CP Violation}\label{sec:Jarlskog}

The quark Yukawa couplings are the only source of CP-violation in the Lagrangian of the SM. The quark Yukawa couplings are described by two complex $3\times3$~matrices $Y_{u}$ and $Y_{d}$ for the up and down quarks. The quark mass matrices are expressed by the Yukawa couplings in the following way
\begin{equation*}
M=\frac{v}{\sqrt{2}}Y_{u},\quad M'=\frac{v}{\sqrt{2}}Y_{d},
\end{equation*}
where $v$ is the vacuum expectation value of the Higgs field and
$M$ and $M'$ are mass matrices of the up- and down-quarks, respectively. Quark running masses are the eigenvalues of the the quark mass matrices and the Cabibbo-Kobayashi-Maskawa matrix  $V$ is obtained from the left bi-unitary dia\-gonalizing matrices $U_{L}^{u,d}$
\begin{gather*}
U_{R}^{u}M{U_{L}^{u}}^{\dagger}=\operatorname{Diag}(m_{t},m_{c},m_{u}),\\
U_{R}^{d}M'{U_{L}^{d}}^{\dagger}=\operatorname{Diag}(m_{b},m_{s},m_{d}),\\
V=U_{L}^{u}{U_{L}^{d}}^{\dagger}.
\end{gather*}
The matrices $U_{L}^{u,d}$ are also obtained from the diagonalization of the Hermitian matrices $Y_{u}^{\dagger}Y_{u}$ and $Y_{d}^{\dagger}Y_{d}$ and we have
\begin{gather*}
U_{L}^{u}Y_{u}^{\dagger}Y_{u}{U_{L}^{u}}^{\dagger}=\operatorname{Diag}(y^{2}_{t},y^{2}_{c},y^{2}_{u}),\\
U_{L}^{d}Y_{d}^{\dagger}Y_{d}{U_{L}^{d}}^{\dagger}=\operatorname{Diag}(y^{2}_{b},y^{2}_{s},y^{2}_{d}),
\end{gather*}
where $y_{t}$, $y_{c}$, $y_{u}$, $y_{b}$, $y_{s}$, $y_{d}$ are the eigenvalues of the quark Yukawa couplings, corresponding to the top, charm, up, bottom, strange and down quarks.

The condition for the presence of CP violation given by Jarlskog in Eq.~\eqref{eq:1} is equivalent to the one given in terms of the matrix~$C$, which is the commutator constructed from the quark Yukawa couplings
\begin{equation}\label{eq:5}
iC=[Y_{u}^{\dagger}Y_{u},Y_{d}^{\dagger}Y_{d}],\quad \det C\neq0 \text{ for CP violation}.
\end{equation}
The determinant $\det C$ is equal
\begin{multline}
\label{eq:6}
\det C=-2 (y^{2}_{t}-y^{2}_{c})(y^{2}_{c}-y^{2}_{u})(y^{2}_{u}-y^{2}_{t})\\
\times(y^{2}_{b}-y^{2}_{s})(y^{2}_{s}-y^{2}_{d})(y^{2}_{d}-y^{2}_{b})J
\end{multline}
and $J$ is the Jarlskog rephasing invariant of the CKM matrix defined by
\begin{equation}
\label{eq:7}
\operatorname{Im}[V_{ij}V_{kl}V^{*}_{il}V^{*}_{kj}]=J\sum_{m,n} \varepsilon_{ikm}\varepsilon_{jln}.
\end{equation}
Thus the CP violation is present in the SM if $J\neq0$ and the quark masses in the up and down sectors are not equal.

\begin{figure}[h]
\centering
\includegraphics[width=\linewidth]{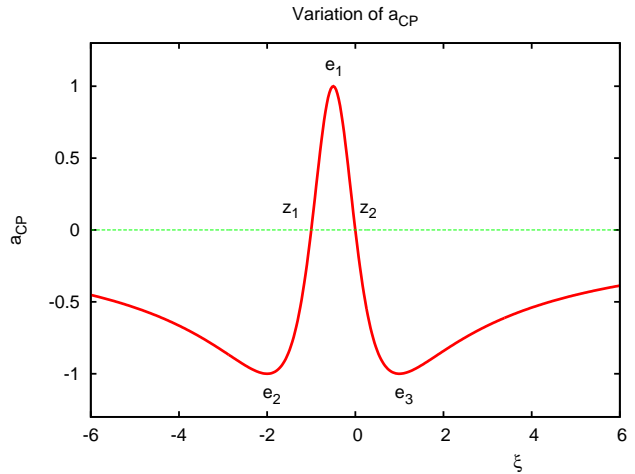}
\caption{The dependence of $a_{CP}$ on $\xi$. The values of $a_{CP}$ are contained in the range $[-1,1]$, and $a_{CP}\rightarrow 0$ for $\vert\xi\vert\rightarrow\infty$.} \label{fig:1}
\end{figure}

Let us now analyze the properties of the matrix $C$. From definition~\eqref{eq:5} it follows that $C$ is Hermitian and traceless. The roots of the characteristic polynomial $w(r)$ of $C$
\begin{equation}
\label{eq:8}
\begin{split}
w(r)=&\det(C-I\cdot r)=-r^{3}+A_{2}r^{2}-A_{1}r+A_{0},\\
&A_{2}=\operatorname{Tr}C=0,\quad A_{1}=-\dfrac{1}{2}\operatorname{Tr}(C^{2}),\\
&A_{0}=\det C=\dfrac{1}{3}\operatorname{Tr}(C^{3})
\end{split}
\end{equation}
are the eigenvalues of $C$ and since $C$ is Hermitian so the eigenvalues must be real. From this we obtain the following condition for the coefficients~$A_{1}$ and~$A_{0}$
\begin{equation}
\label{eq:9}
\begin{split}
a_{CP}=-\dfrac{3\sqrt{3}A_{0}}{2A_{1}\sqrt{-A_{1}}} &=\dfrac{3\sqrt{6}\det C}{(\sqrt{\operatorname{Tr}(C^{2})})^{3}}
=\dfrac{\sqrt{6}\operatorname{Tr}(C^{3})}{(\sqrt{\operatorname{Tr}(C^{2})})^{3}},\\
-1&\leq a_{CP}\leq 1.
\end{split}
\end{equation}
The parameter $a_{CP}$ and the inequality in~\eqref{eq:9} were introduced by Jarlskog~\cite{PhysRevD.36.2128,*jarlskog1989cp}. From definition~\eqref{eq:9} one can see that $a_{CP}$ depends only on the eigenvalues $(r_{1},r_{2},r_{3})$ of the matrix $C$
\begin{equation}
\label{eq:10}
a_{CP}=\dfrac{3\sqrt{3}r_{1}r_{2}r_{3}}{2(-(r_{1}r_{2}+r_{1}r_{3}+r_{2}r_{3}))^{\frac{3}{2}}}.
\end{equation}
Moreover, using the condition $\operatorname{Tr}C=r_{1}+r_{2}+r_{3}=0$ and dividing the numerator and denominator of Eq.~\eqref{eq:10} by $r_{3}^{3}$ we find that $a_{CP}$ is a function of only one parameter
\begin{equation}
\label{eq:11}
a_{CP}=-\dfrac{3\sqrt{3}\xi(1+\xi)}{2(1+\xi+\xi^{2})^{\frac{3}{2}}}, \quad\xi=\dfrac{r_{1}}{r_{3}}.
\end{equation}
In Fig.~\ref{fig:1} we draw the dependence of $a_{CP}$ on $\xi$. One can see that $a_{CP}$ has
\noindent one maximum, two minima and two zeros. They are marked in the figure by the letters $e_{i}$ and $z_{i}$ and their meaning is given in Table~\ref{table:2}.
\begin{table}[h]
\centering
\begin{tabular}{r|cll}
&$\xi$&Type&Eigenvalues\\
\hline
$e_{1}$&-0.5&maximum&$(r,r,-2r)$\\
$e_{2}$&-2.0&minimum&$(-2r,r,r)$\\
$e_{3}$&+1.0&minimum&$(r,-2r,r)$\\
$z_{1}$&-1.0&zero&$(r,0,-r)$\\
$z_{2}$&\phantom{+}0.0&zero&$(0,r,-r)$
\end{tabular}
\caption{Extrema and zeros of $a_{CP}$}
\label{table:2}\centering
\end{table}
From this table we see that the value of $a_{CP}$ (which is invariant under the rescaling of the eigenvalues) characterizes the relative distribution of the eigenvalues of the matrix $C$. The structure of the eigenvalues at the minimum and the two maxima is the same and is $(r,r,-2r)$ and at zeros it is $(r,-r,0)$

We conclude our discussion of the Jarlskog's description of CP violation by listing the observables sensitive to the CP violation built from the $C$ matrix
\begin{enumerate}
\item $\det C=\dfrac{1}{3}\operatorname{Tr}(C^{3})$.
\item $\operatorname{Tr}(C^{2})$.
\item $a_{CP}=\dfrac{\sqrt{6}\operatorname{Tr}(C^{3})} {(\sqrt{\operatorname{Tr}(C^{2})})^{3}}.$
\item Jarlskog's phase invariant $J$.
\item Eigenvalues of the $C$ matrix.
\end{enumerate}
We will analyze the renormalization group evolution of these observables to determine the properties of the CP violation at the Planck scale.

\section{Renormalization Group Evolution\\ of CP-violation}\label{sec:CP-evolution}

The renormalization group evolutions for our observables are obtained from the RGE for the parameters of the SM that are listed in Table~\ref{table:1}. Let us write the RGE for these parameters in the generic form~\footnote{This form of the RGE equations is valid for any number of loops with the corresponding $\beta$ functions.}
\begin{equation}\label{eq:12}
\begin{aligned}
&\dfrac{dg_{i}}{dt}=g_{i}^{3}\tilde{\beta}_{g_{i}},&
&i=1,2,3,&
&\dfrac{dY_{u}}{dt}=Y_{u}\tilde{\beta}_{Y_{u}},\\
&\dfrac{dY_{d}}{dt}=Y_{d}\tilde{\beta}_{Y_{d}},&
&\dfrac{dY_{e}}{dt}=Y_{e}\tilde{\beta}_{Y_{e}},&
&\dfrac{d\lambda}{dt}=\beta_{\lambda}.
\end{aligned}
\end{equation}
Here $\tilde{\beta}_{g_{i}}$ and $\beta_{\lambda} $ are scalar functions and $\tilde{\beta}_{Y_{u}} $, $\tilde{\beta}_{Y_{d}}$ and $\tilde{\beta}_{Y_{e}} $ are $ 3\times3 $ Hermitian matrices that \emph{do not commute} with $ Y_{u} $, $ Y_{d} $ and $ Y_{e} $. The $\tilde{\beta}$ functions are polynomials of $g_{i}$, $\lambda$, $Y_{u}^{\dagger}Y_{u}$, $Y_{d}^{\dagger}Y_{d}$ and $Y_{e}^{\dagger}Y_{e}$. The power of the polynomials increases with the number of loops. The RGE in the SM are thus a set of coupled, non linear ordinary differential equations. The exact solution of these equations is in general not possible.

We are interested in the evolution of the CP violating observables discussed in the previous section. All these observables are obtained from the $C$ matrix. The RGE for the matrix $C$ can be obtained from Eqs.~\eqref{eq:12} and it is equal
\begin{multline}
\label{eq:13}
\dfrac{dC}{dt}=\{\tilde{\beta}_{Y_{u}}+\tilde{\beta}_{Y_{d}},C\}-
i\{[\tilde{\beta}_{Y_{u}},Y_{d}^{\dagger}Y_{d}],Y_{u}^{\dagger}Y_{u}\}\\ -i\{[Y_{u}^{\dagger}Y_{u},\tilde{\beta}_{Y_{d}}],Y_{d}^{\dagger}Y_{d}\}.
\end{multline}
The right hand side of this equation depends on $g_{i}$, $\lambda$, $Y_{u}^{\dagger}Y_{u}$, $Y_{d}^{\dagger}Y_{d}$, so it requires that the solutions of Eqs.~\eqref{eq:12} are known. From the solution of Eqs.~\eqref{eq:12} one can compute the matrix $C$ directly, so we will not use Eq.~\eqref{eq:13} in our analysis and we will numerically solve Eqs.~\eqref{eq:12}.

To solve Eqs.~\eqref{eq:12} we need the initial values, for which we choose the representation in which $Y_{d}(0)$ is diagonal and $Y_{u}(0)$ 
contains the CKM matrix
\begin{align}
Y_{u}(0)&=\operatorname{Diag}(y_{t},y_{c},y_{u})\cdot V_{\text{CKM}},\label{eq:14}\\
Y_{d}(0)&=\operatorname{Diag}(y_{b},y_{s},y_{d}).\label{eq:15}
\end{align}
The initial values of all the parameters are taken from the PDG Book of Particle Properties~\cite{PhysRevD.86.010001}.

The initial values of the CKM matrix and the eigenvalues of the quark and lepton Yukawa couplings exhibit a strong hierarchy, which may have influence on the precision of the numerical analysis. To confirm the numerical calculations we have calculated the leading terms for the $C$ matrix and $\det(C)$
\begin{equation}\label{eq:16}
\begin{aligned}
C_{12}& \sim -i y_{t}^{2}y_{b}^{2}V_{11}V_{21}^{*}&
C_{13}& \sim -i y_{t}^{2}y_{b}^{2}V_{11}V_{31}^{*}\\
C_{23}& \sim -i y_{c}^{2}y_{b}^{2}V_{21}V_{31}^{*}\quad&
\det(C)& \sim -2y_{t}^{4}y_{b}^{4}y_{c}^{2}y_{s}^{2}J
\end{aligned}
\end{equation}
and we found that the results based on approximate formulas agree with very high precision with the numerical calculations.

\section{Discussion of the Results}\label{sec:discussion}

In this section we will present the numerical results of the study of all characteristic parameters of the matrix $C$ that were analyzed in the energy range from $m_{t}$ to $10^{16}$~GeV.

\begin{figure}[h]
\centering
\includegraphics[width=\linewidth]{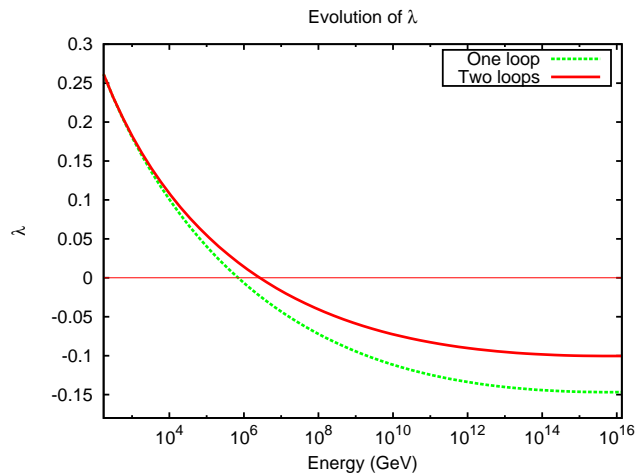}
\caption{The dependence of the Higgs quartic coupling $\lambda$ on energy. Only the region of energy where $\lambda$ is positive has stable vacuum and is physically acceptable. The top mass is equal $m_{t}=173.07$. The difference for one and loop evolution is significant.} \label{fig:2}
\end{figure}

The renormalization group method is the tool to determine the range of the validity of the SM. We will start by showing the evolution of the Higgs quartic coupling~$\lambda$, which must be positive for a stable theory. The evolution of $\lambda$ is shown in Fig.~\ref{fig:2} in which we see that around $ 10^{6}-10^{7} $~GeV $\lambda$ becomes negative and the threshold of new physics should be below this energy, when considering the observed value for $m_{t}$~\footnote{The exhaustive discussion of the vacuum stability is contained in a recent paper~\cite{espinosa2013Mainz}. One should notice that the value of the energy, where $\lambda$ becomes negative is lower in our case than in Ref.~\cite{espinosa2013Mainz}. This may be due to a different approximation, used in~\cite{espinosa2013Mainz} than in our analysis. We do not neglect any terms in our computations.}.
\begin{figure}[h]
\centering
\includegraphics[width=\linewidth]{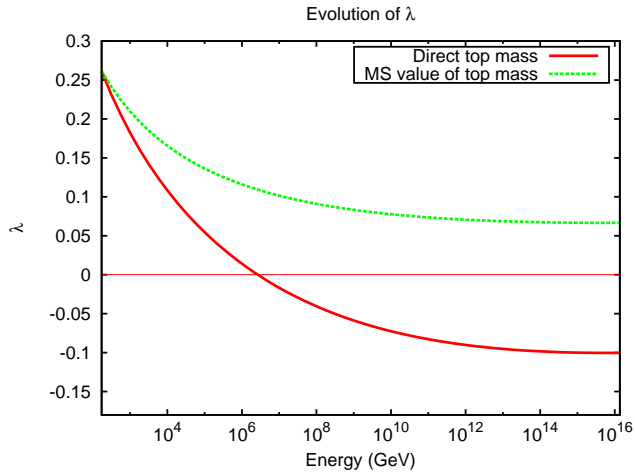}
\caption{The dependence of the Higgs quartic coupling $\lambda$ on energy for two values of top quark mass: directly measured mass  $m_{t}=173.07$~GeV and $\overline{\text{MS}}$ mass $m_{t}=160$~GeV. Here we include only two loop evolution of $\lambda$.} \label{fig:2n}
\end{figure}

The evolution of $\lambda$ depends strongly on the value of the top quark mass. In Fig.~\ref{fig:2} the top quark mass was taken from~\cite{PhysRevD.86.010001} as the directly measured mass, which is equal $m_{t}=173.07$~GeV (we do not show the influence of the top quark mass experimental errors in the evolution). In Fig.~\ref{fig:2} we compare the one loop and two loop evolutions. The difference in the evolutions is significant and for more precise predictions one should use the two loop evolution.

Particle Data Group~\cite{PhysRevD.86.010001} also quotes the $\overline{\text{MS}}$ top quark mass from cross section measurements which is
equal $m_{t}=160\genfrac{}{}{0pt}{1}{+5}{-4}$~GeV. In Fig.~\ref{fig:2n} we show the two loop $\lambda$ evolution for the two values of the top quark mass: $m_{t}=160$~GeV and $m_{t}=173.07$~GeV. We notice that for the  $\overline{\text{MS}}$ top quark mass the quartic coupling $\lambda$ is positive for the whole range of energy up to the Planck scale. This fact may suggest that the range of validity of the SM may be larger than the one obtained from the directly measured top quark mass.

\begin{figure}[h]
\centering
\includegraphics[width=\linewidth]{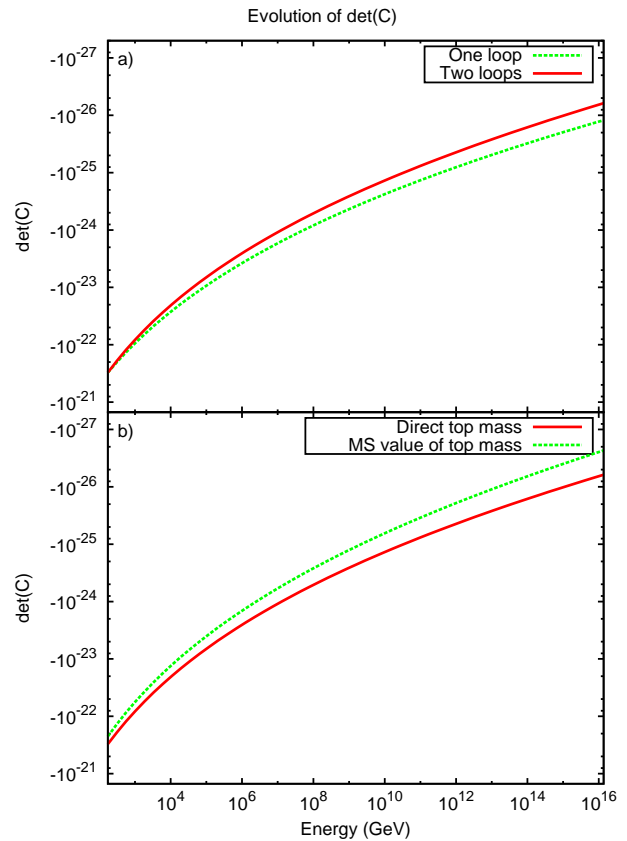}
\caption{RG evolution of $\det C$. In a) we show the dependence of $\det C$ on the number of loops and in b) we show the two loop evolution of $\det C$ for two values of the top quark mass: $m_{t}=173.07$~GeV and $m_{t}=160$~GeV.} \label{fig:3}
\end{figure}

Let us now start the discussion of the evolution of the CP-violation observables with the $\det C$. In Fig.~\ref{fig:3} we display the RGE evolution of $\det C$ for one and the two loop cases and for the top quark masses $m_{t}=160$~GeV and $m_{t}=173.07$~GeV.  The variation of $\det C$ is very significant, because its absolute value is reduced by~5~orders of magnitude in the considered range of energies. The non vanishing of $\det C$ is the criterion of CP~violation in the SM, so such a dramatic reduction of $\det C$ might indicate that at the unification energy the rate of CP~violation in the SM is also reduced. From Eq.~\eqref{eq:6} we know that $\det C$ is the product of the Jarlskog's~$J$ and of the quark mass differences, so we must study other parameters of the $C$~matrix to be able to interpret the evolution of $\det C$. The other information contained in Fig.~\ref{fig:3} 
is the dependence of $\det C$ on the number of loops and on the top quark mass. The dependence of the evolution of $\det C$ on these two parameters is well marked, but the overall physical picture does not depend strongly on these parameters: the absolute value of $\det C$ is decreasing by 5~orders of magnitude between the top mass and the Planck scale.

\begin{figure}[h]
\centering
\includegraphics[width=\linewidth]{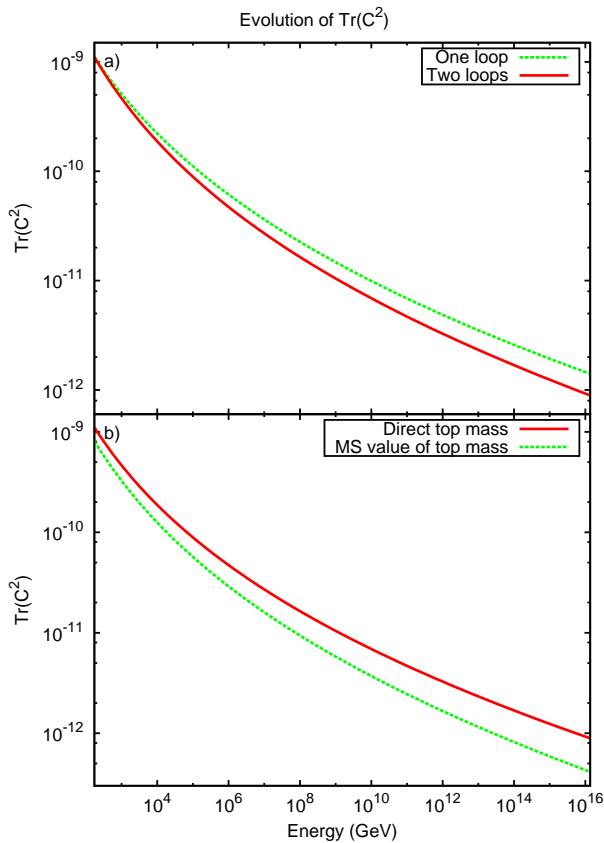}
\caption{RG evolution of $\operatorname{Tr}C^{2}$. In a) we show the dependence of  $\operatorname{Tr}C^{2}$ on the number of loops and in b) we show the two loop evolution of  $\operatorname{Tr}C^{2}$ for two values of the top quark mass: $m_{t}=173.07$~GeV and $m_{t}=160$~GeV.}\label{fig:4}
\end{figure}

The next important parameter of the matrix $C$ is  $\operatorname{Tr}C^{2}$, which as $\det C$ is the rephasing invariant of the quark fields, but it does not vanish, when CP is conserved, but it is used in the ratio with $\det C$ to determine the normalized rate of CP violation. In Fig.~\ref{fig:4} we show the evolution of $\operatorname{Tr}C^{2}$, whose variation is also very significant. The value of $\operatorname{Tr}C^{2}$ is reduced by 3~orders of magnitude in the considered energy range. The explicit formula for $\operatorname{Tr}C^{2}$ in terms of quark masses and the CKM matrix is rather complicated, so there is no simple interpretation of such an evolution in terms of other observables. The dependence of $\operatorname{Tr}C^{2}$ on the number of loops and on the top quark mass is also well marked, but again the physical picture is similar in all cases.

\begin{figure}[h]
\centering
\includegraphics[width=\linewidth]{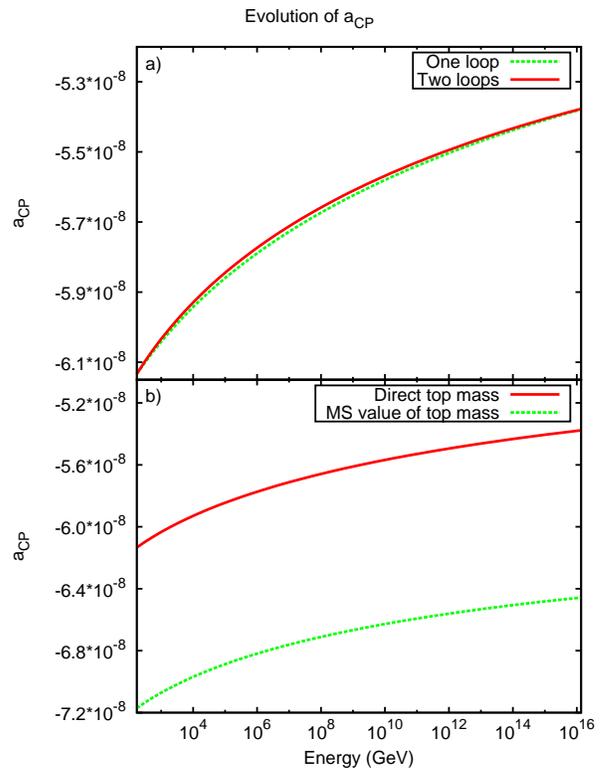}
\caption{RG evolution of $a_{CP}$.  In a) we show the dependence of  $a_{CP}$ on the number of loops and in b) we show the two loop evolution of  $a_{CP}$ for two values of the top quark mass: $m_{t}=173.07$~GeV and $m_{t}=160$~GeV.}\label{fig:5}
\end{figure}

Another important parameter to study is $a_{CP}$,  defined in~\eqref{eq:9}, which is the ratio of $\det C$ and $(\operatorname{Tr}C^{2})^{(3/2)}$. In Fig~\ref{fig:5} we show the evolution of $a_{CP}$. The parameter $a_{CP}$ is invariant upon the rescaling of the Yukawa couplings (and quark masses). From Fig.~\ref{fig:5} we see that $a_{CP}$ has rather slow energy dependence and the fast evolution of $\det C$ and  $\operatorname{Tr}C^{2}$ mostly cancel each other in $a_{CP}$. Note that the scale in Fig.~\ref{fig:5} is linear while in Figs.~\ref{fig:3} and~\ref{fig:4} it is logarithmic. 
The overall change of $a_{CP}$ is approximately 6~\%, but the difference 
between the one and two loop evolution is very small. The initial value of $a_{CP}$ for the top quark masses $m_{t}=173.07$~GeV and $m_{t}=160$~GeV differ, but the pattern of the evolution for both quark masses is similar.

\begin{figure}[h]
\centering
\includegraphics[width=\linewidth]{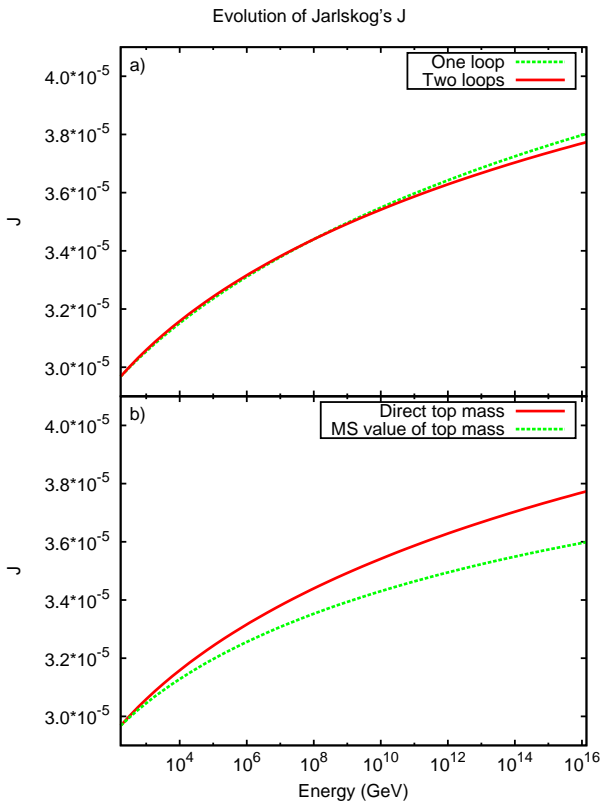}
\caption{RG evolution of the Jarlskog's parameter $J$.  In a) we show the dependence of  $J$ on the number of loops and in b) we show the two loop evolution of ~$J$ for two values of the top quark mass: $m_{t}=173.07$~GeV and $m_{t}=160$~GeV.}\label{fig:6}
\end{figure}

From Fig.~\ref{fig:5} we also notice that the value of the dimensionless parameter $a_{CP}$ is of the order $10^{-10}$ and it is \textit{very} small. From the analysis of the parameter $a_{CP}$ \mbox{in Sec.~\ref{sec:Jarlskog}} we deduce that the scenario for the eigenvalues of the matrix~$C$ that is realized is the one that is close to the zero value of $a_{CP}$, i.e., there are two eigenvalues that almost cancel each other and the third eigenvalue is much smaller, than the remaining two.
In Fig.~\ref{fig:6} we show the evolution of the Jarlskog's phase invariant~$J$. We see that $J$ grows with the energy and the overall increase of $J$ up to the Planck's scale is approximately 30~\%. The two loop modification of $J$ is very small. The growth of~$J$ for the $\overline{\text{MS}}$ top quark mass $m_{t}=160$~GeV is slower than for the directly measured $m_{t}=173.07$~GeV.
\begin{figure}[h]
\centering
\includegraphics[width=\linewidth]{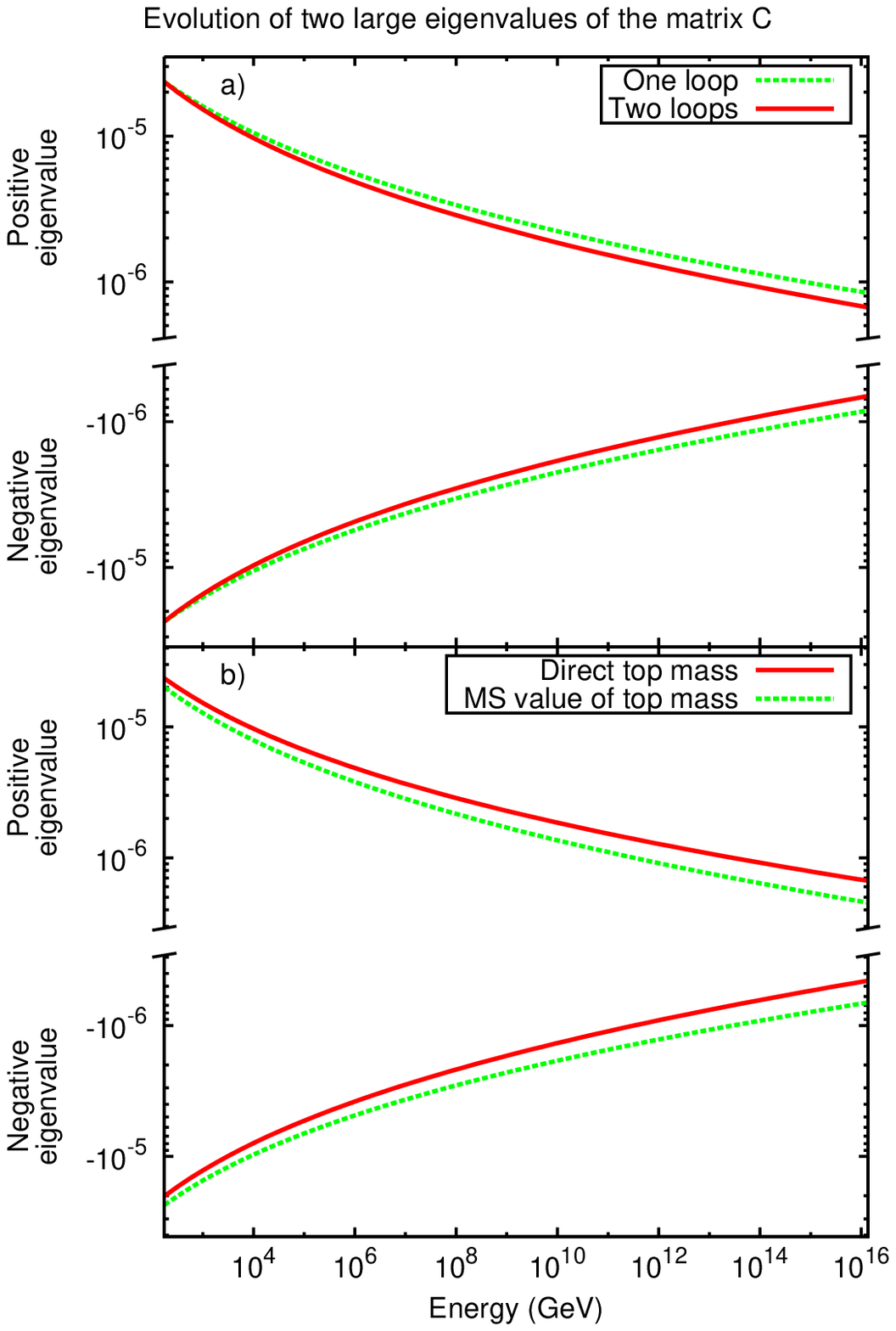}
\caption{RG evolution of the two \textit{large} eigenvalues of the matrix $C$.  In a) we show the dependence of the eigenvalues on the number of loops and in b) we show the two loop evolution of the eigenvalues for two values of the top quark mass: $m_{t}=173.07$~GeV and $m_{t}=160$~GeV.}\label{fig:7}
\end{figure}
\begin{figure}[h]
\centering
\includegraphics[width=\linewidth]{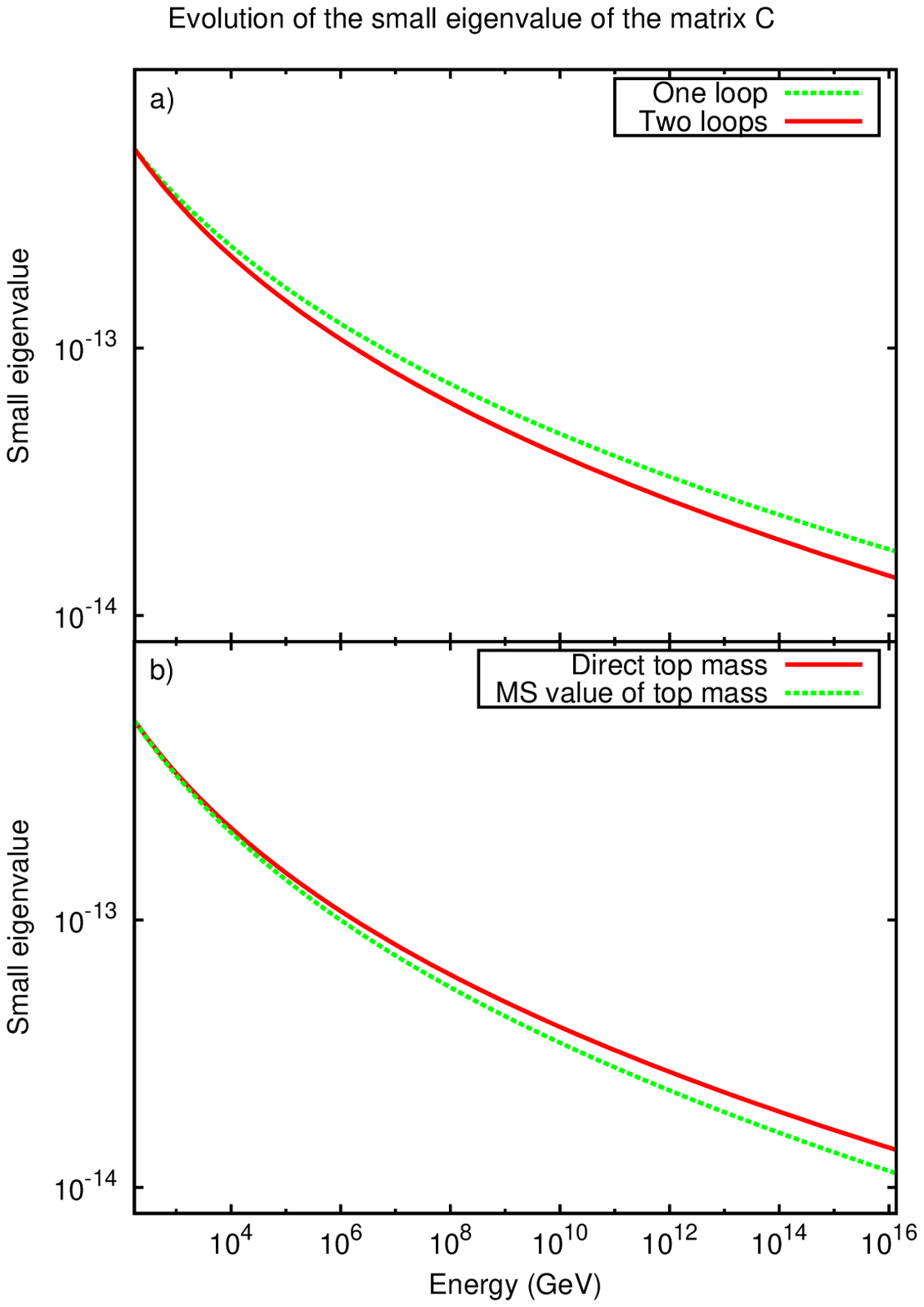}
\caption{RG evolution of the \textit{small} eigenvalue of the matrix $C$.  In a) we show the dependence of the eigenvalue on the number of loops and in b) we show the two loop evolution of the eigenvalue for two values of the top quark mass: $m_{t}=173.07$~GeV and $m_{t}=160$~GeV.}\label{fig:8}
\end{figure}

Finally in Figs.~\ref{fig:7} and~\ref{fig:8} we show the evolution of the eigenvalues of the matrix~$C$. The structure of the evolution of the eigenvalues is consistent with the discussion of the evolution of $a_{CP}$, $\det C$ and $\operatorname{Tr}C^{2}$:
\begin{enumerate}	
\item From the small value of $a_{CP}$ it follows that there are two \textit{large} and one \textit{small} eigenvalue of the matrix~$C$.
\item The absolute values of the eigenvalues are quickly decreasing with energy.
\end{enumerate}
The dependence of the evolution on the number of loops and on the top quark mass is important, but does not bring in new physical effects.

\section{Conclusions}\label{sec:conclusions}

Let us start our conclusions with the analysis of the range of validity of the SM. The sign of the Higgs quartic coupling is the criterion of the vacuum stability. For negative values of $\lambda$ the model is unstable. In Figs.~\ref{fig:2} and ~\ref{fig:2n} we compare the evolution of $\lambda$ for one and two loops renormalization group equations and for two different values of the top quark mass reported by PDG~\cite{PhysRevD.86.010001}. From Fig.~\ref{fig:2} one can see that the evolution of $\lambda$ is sensitive to the number of loops and the difference at the Planck scale for the two cases is significant. The dependence of the evolution of $\lambda$ on the top quark mass is very strong. In Fig.~\ref{fig:2n} we see that for the $\overline{\text{MS}}$ top quark mass $m_{t}=160$~GeV the quartic coupling $\lambda$ is positive in the whole energy range up to the Planck mass. On the other hand for the directly measured quark mass $m_{t}=173.07$ the model becomes unstable at $\sim10^{7}$~GeV. In our opinion the results of the renormalization group analysis of the model based on the $\overline{\text{MS}}$ top quark mass should be taken on the equal footing with the ones obtained from the directly measured top quark mass. The subject of the stability of vacuum has been extensively discussed in the recent paper~\cite{espinosa2013Mainz} and we will not delve into it further, because it is not the main subject of our paper. We just conclude that it is justified to consider the renormalization group evolution of the SM up to the Planck energy.

The renormalization group evolution of the CP-observables has two patterns: vary fast variation of $\det C$ and $\operatorname{Tr}C^{2}$ and relatively slow dependence on energy of Jarlskog's $J$ and $a_{CP}$.
The absolute values of $\det C$ and $\operatorname{Tr}C^{2}$ \textit{decrease} with the energy 5 orders and 3 orders of magnitude, respectively. Jarlskog's $J$ and $a_{CP}$ are invariant upon rescaling of the quark Yukawa couplings, because they are the ratios of $\det C$ and the suitable powers of the eigenvalues of the quark Yukawa couplings. These parameters have slow variation with energy and one can conclude that the diagonalizing matrices of the Yukawa couplings and the CKM matrix also have slow dependence on energy. The evolution of $J$ is very remarkable: the one and two loop evolution are almost identical, but the dependence of $J$ on the top quark mass is significant (see Fig.\ref{fig:6}). The coefficient $J$ grows with the energy and its value at the Planck's scale is approximately 25~\% larger than at the top mass. Such a value is not sufficient for the explanation of the cosmological analysis of the baryon asymmetry and new sources of CP-violation are needed (see, e.g., a recent paper on a discussion of the two Higgs doublets extension of the SM, compatible with the observed value of the Higgs boson,~\cite{PhysRevLett.111.091801}).

The next important result is the structure of the eigenvalues of the $C$ matrix. If one expands the matrix $C$ in powers of the CKM matrix $\lambda_{\text{CKM}}$, then the leading order term of $\det C$ vanishes. The approximate values of the eigenvalues of the $C$ matrix are equal 
\begin{equation}
\label{eq:17}
(y_{t}^{2}y_{b}^{2}\lambda_{\text{CKM}},-y_{t}^{2}y_{b}^{2}\lambda_{\text{CKM}},2y_{c}^{2}y_{s}^{2}\lambda_{\text{CKM}}^{4}\eta_{\text{CKM}}),
\end{equation}
where $\lambda_{\text{CKM}}$ and $\eta_{\text{CKM}}$ are the parameters of the CKM matrix in the Wolfenstein parameterization. Notice that the first two eigenvalues cancel each other exactly in this approximation. The smallness of $a_{CP}$ follows from the smallness of the third eigenvalue of the matrix $C$ and this is a consequence of the hierarchy of the quark Yukawa couplings. Notice, however, that $a_{CP}$ would also be zero if any  of the two Yukawa couplings in the up or down quark sectors were equal.

\begin{acknowledgments}
S.R.J.W. acknowledges partial support from projects SIP$20130588$-IPN and SIP$20140693$-IPN, also to EDI and Comisión de Operación y Fomento de Actividades Académicas (COFAA-IPN). 
\end{acknowledgments}


%

\end{document}